\documentstyle[prl,aps,preprint,epsf]{revtex}

\begin{document}

\noindent{\bf Ricci-Tersenghi et al. reply:} In their
comment~\cite{COMMENT} to our paper~\cite{US} de Candia and Coniglio
show evidence that the equilibrium overlap distribution $P(q)$ of the 
frustrated Ising lattice gas (FILG), do not coincide with the one that 
could be estimated from the off-equilibrium correlation and response functions
presented in~\cite{US}, for low temperatures and large chemical potentials.  
From the theoretical point of view one expects the
relation between static and dynamic quantities to hold as long as
one-time observables (like energy or density) converge to the
equilibrium values~\cite{FRAMEPAPE}.

In order to clarify what is happening in the glassy phase of the FILG,
we have performed a new large set of simulations, working mainly at
$\beta J = \infty$ and $\beta \mu = 10$.  At equilibrium we find a
$P(q)$ with two clear peaks and a continuous part in between (like
in~\cite{COMMENT}). Whether the continuous part disappears in the
thermodynamical limit, thus suggesting a one step replica symmetry
broken (RSB) solution, is hard to decide at present.

\begin{figure}
\epsfxsize=\columnwidth
\epsffile{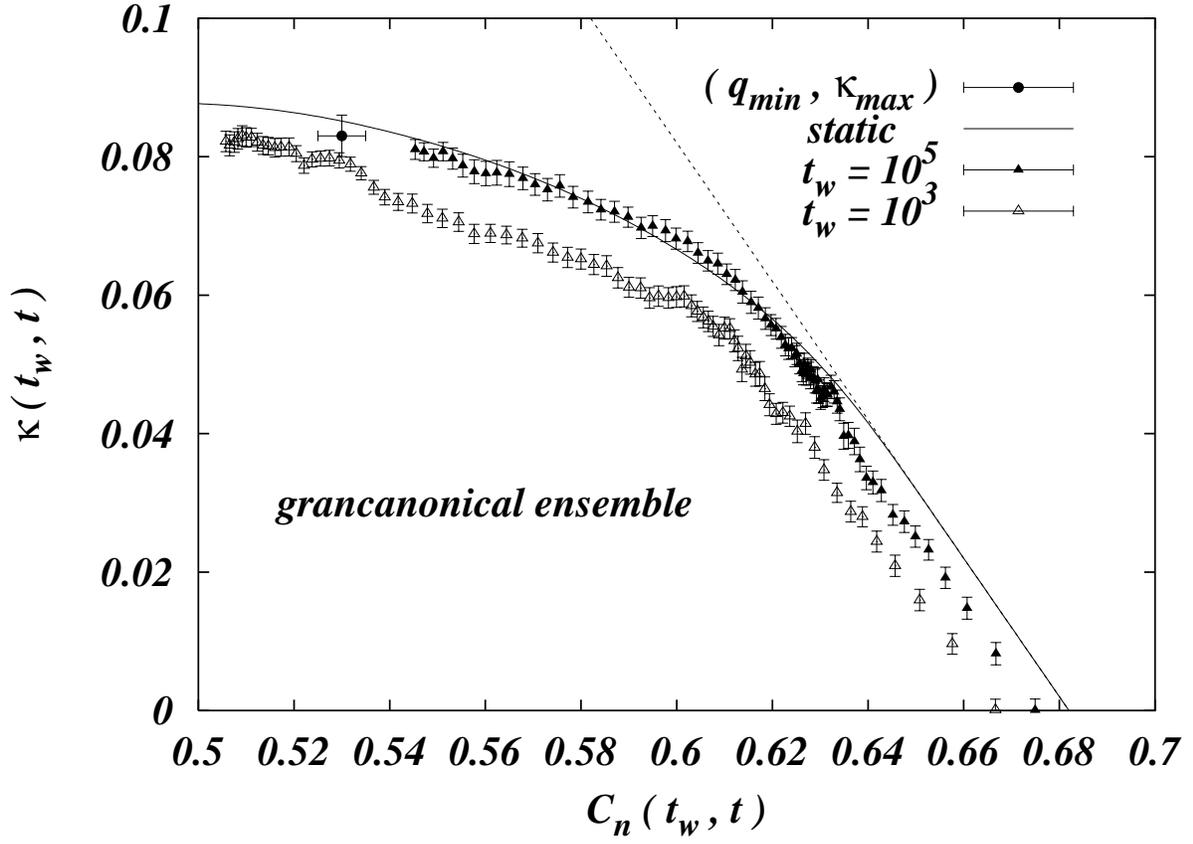}
\caption{Grancanonical response versus correlation.}
\label{fig1}
\end{figure}

In the out of equilibrium regime (quenching from $\beta \mu = -\infty$
to $\beta \mu = 10$) we have repeated all the experiments presented
in~\cite{US}, but now letting the density grow even after the waiting
time $t_w$.  The results (see Fig.~\ref{fig1}) are indeed different
from those where the density is kept fixed after time $t_w$.  In
Fig.~\ref{fig1} we clearly see that for large waiting times the
agreement between static and dynamic susceptibilities is very good
indicating that the system is not trapped in any long-lived metastable
state.

\begin{figure}
\epsfxsize=\columnwidth
\epsffile{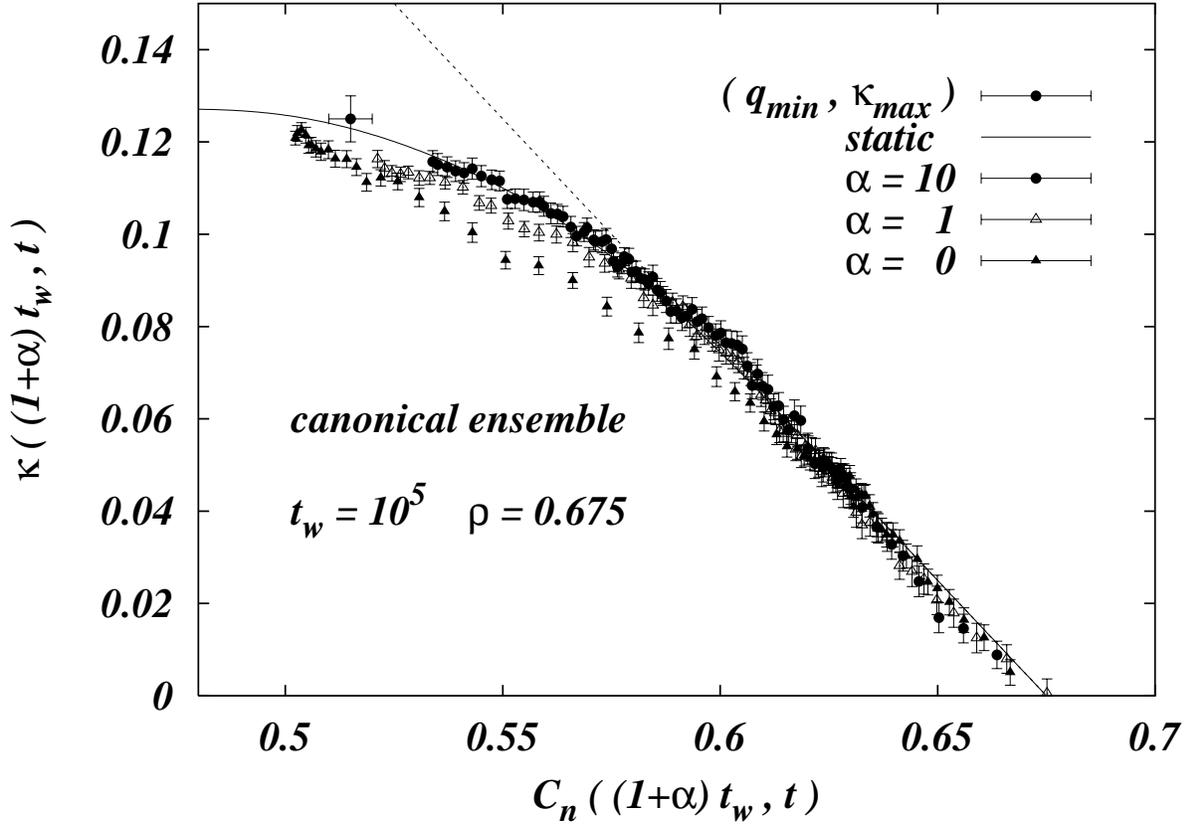}
\caption{Canonical response versus correlation.}
\label{fig2}
\end{figure}

Still the open question is why in the experiments presented
in~\cite{US} (and confirmed in~\cite{COMMENT}) the responses are so
different and seem not to be related to the corresponding static
susceptibilities.  The answer is related to the sudden change of
dynamics at time $t_w$: up to $t_w$ we evolved the system in a
grancanonical ensemble where the density tends to increase thanks to
the large chemical potential, while after time $t_w$ we kept the
density fixed, allowing the system evolve only in a canonical
ensemble.  In order to verify this hypothesis we have done the
following numerical experiment: after quenching the system we let it
evolve for $t_w$ time steps within the grancanonical ensemble and for
$\alpha \, t_w$ more in the canonical one.  Finally at time
$(1+\alpha) t_w$ we switch the field on and we measure the response
[always with a fixed density $\rho(t_w)$].  For $\alpha=0$ we recover
the behavior reported in~\cite{COMMENT,US}.  Nevertheless for large
$\alpha$ the system ``forgets'' the drastic change of the dynamics
done at time $t_w$, thermalizing in the canonical ensemble at a fixed
density $\rho(t_w)$ and the dynamical susceptibilities are in
agreement with the static ones measured at the same density as can be
seen in Fig.~\ref{fig2}.

It must be emphasized that, in both figures, the off-equilibrium
results obtained with $L=40$ systems are still compatible with linear
fits and that the curvature of the static result ($L=10$) may show
strong finite size effects.

Concluding, we have shown that the link between static and dynamic
susceptibilities is valid on the time scales we can reach in our
simulations for both the canonical and the grancanonical ensembles.
Regarding the kind of RSB solution present in the glassy phase of the
3d FILG (whether it is characterized by one or infinite RSB steps), we
believe that further studies are needed in order to clarify this
point.  The model still seems to be a good candidate for a 3d system
with 1-RSB.

\noindent
F. Ricci-Tersenghi$^1$, G. Parisi$^2$, D.A. Stariolo$^3$ and
J.J. Arenzon$^3$

\noindent{\small $^1$ Abdus Salam ICTP, Trieste (Italy)\\
$^2$ Universit\`a di Roma ``La Sapienza'', Roma (Italy)\\
$^3$ Univ. Fed. Rio Grande do Sul, Porto Alegre RS (Brazil)}

\end{document}